**Magnetic resonance delta radiomics to track radiation response in lung tumors receiving stereotactic MRI-guided radiotherapy**

**Authors:** Yining Zha, BS[1,2,3+]; Benjamin H. Kann, MD[1,2+*]; Zezhong Ye, PhD[1,2]; Anna Zapaishchykova, MS[1,2,4]; John He, BA[2]; Shu-Hui Hsu PhD[2]; Jonathan E. Leeman, MD[2]; Kelly J. Fitzgerald, MD, PhD[2]; David E. Kozono, MD, PhD[2]; Raymond H. Mak, MD[1,2]; Hugo J.W.L. Aerts, PhD[1,2,4,5]*

**Affiliations:**

1. Artificial Intelligence in Medicine Program, Mass General Brigham, Harvard Medical School, Boston, MA, USA
2. Department of Radiation Oncology, Dana-Farber Cancer Institute and Brigham and Women's Hospital, Harvard Medical School, Boston, MA, USA
3. Department of Biostatistics, Harvard T.H. Chan School of Public Health, Boston, MA, USA
4. Radiology and Nuclear Medicine, CARIM & GROW, Maastricht University, Maastricht, the Netherlands
5. Department of Radiology, Brigham and Women's Hospital, Dana-Farber Cancer Institute, Harvard Medical School, Boston, MA, USA

* Corresponding author

+ These authors contributed equally to this manuscript.

**Correspondence address to:**

Benjamin H. Kann, M.D.
Department of Radiation Oncology,
Dana-Farber Cancer Institute and Brigham and Women's Hospital,
Harvard Medical School, 75 Francis Street, Boston, MA 02115, MA, USA
Tel: +1 617-732-6310
Email: Benjamin_Kann@dfci.harvard.edu

Hugo J.W.L. Aerts, PhD
Department of Radiation Oncology,
Dana-Farber Cancer Institute and Brigham and Women's Hospital,
Harvard Medical School, 75 Francis Street, Boston, MA 02115, MA, USA

Tel: + 1 617-525-7156
Email: haerts@bwh.harvard.edu

**Running Title**: Delta-radiomics analysis for lung cancer SBRT MR dose response

**Author Contributions**

Study design: Y.Z., B.H.K.; code design, implementation and execution: Y.Z., B.H.K.; acquisition, analysis or interpretation of data: Y.Z., B.H.K., Z.Y., A.Z., J.H., S.H.H.; statistical analysis: Y.Z., B.H.K.; writing of the manuscript: Y.Z., B.H.K.; critical revision of the manuscript for important intellectual content: all authors; study supervision: B.H.K.

**Data Availability**

Although raw MR imaging data cannot be shared, all measured results to replicate the statistical analysis will be shared once the manuscript is accepted for publication.

**Code Availability**

The codes of the delta-radiomics pipeline and statistical analysis are available at: https://github.com/AIM-KannLab/lung_cancer_radiomics.

**Funding**

The authors acknowledge support from the National Institutes of Health (NIH) with grant numbers (NIH U24CA194354, NIH U01CA190234, NIH U01CA209414, NIH R35CA22052, NIH K08DE030216, NIH 5F31DE031502-02 and NIH 5T32CA261856), the European Union-European Research Council (866504), the Radiological Society of North America (RSCH2017) and the unrestricted grant of Stichting Hanarth Fonds. All analyses and conclusions in this manuscript are the sole responsibility of the authors and do not necessarily reflect the opinions or views of the clinical trial investigators, the NCTN, or the NCI.

**Competing Interests**

The other authors declare no conflict of interests.

## ABSTRACT

**Introduction**: Lung cancer is a leading cause of cancer-related mortality, and stereotactic body radiotherapy (SBRT) has become a standard treatment for early-stage lung cancer. However, the heterogeneous response to radiation at the tumor level poses challenges. Currently, standardized dosage regimens lack adaptation based on individual patient or tumor characteristics. Thus, we explore the potential of delta radiomics from on-treatment magnetic resonance (MR) imaging to track radiation dose response, inform personalized radiotherapy dosing, and predict outcomes.

**Methods**: A retrospective study of 47 MR-guided lung SBRT treatments for 39 patients was conducted. Radiomic features were extracted using Pyradiomics, and stability was evaluated temporally and spatially. Delta radiomics were correlated with radiation dose delivery and assessed for associations with tumor control and survival with Cox regressions.

**Results**: Among 107 features, 49 demonstrated temporal stability, and 57 showed spatial stability. Fifteen stable and non-collinear features were analyzed. Median Skewness and surface to volume ratio decreased with radiation dose fraction delivery, while coarseness and 90th percentile values increased. Skewness had the largest relative median absolute changes (22%-45%) per fraction from baseline and was associated with locoregional failure (p=0.012) by analysis of covariance. Skewness, Elongation, and Flatness were significantly associated with local recurrence-free survival, while tumor diameter and volume were not.

**Conclusions**: Our study establishes the feasibility and stability of delta radiomics analysis for MR-guided lung SBRT. Findings suggest that MR delta radiomics can capture short-term radiographic manifestations of intra-tumoral radiation effect.

## INTRODUCTION

Lung cancer is the leading cause of cancer-related mortality with a 3-year relative survival rate of 33% in the United States from 2016 to 2018 (Siegel et al., 2023). Stereotactic body radiotherapy (SBRT) has emerged as a standard treatment modality for early-stage lung cancer in the inoperable setting with high rates of initial local control but a risk of recurrence that increases with time (Timmerman et al., 2010). Previous studies have demonstrated that pathologic complete response post-SBRT is heterogeneous and lower than anticipated (Palma et al., 2019). These findings suggest there may be differential radiation dose responses at the tumor level, even at ablative dose ranges. Currently, dosage regimens are standardized without adapting doses based on individual patient or tumor characteristics and are focused on reaching a biologically equivalent dose (BED) >= 100 Gy (alpha/beta: 10), with tailored tradeoffs in situations where tumors closely abut critical normal structures (Weykamp et al., 2023). There is currently no direct way to assess radiation dose response during SBRT in vivo, and thus little basis for tailoring dose. Improving the ability to track and predict individual, patient-level responses to SBRT would be valuable to inform personalized radiotherapy (RT) dosing regimens, risk-stratification, and selection of patients for adjuvant or neoadjuvant therapy intensification.

Radiation damage is exerted at the cellular level by DNA double-stranded breaks, which cascade to programmed cell death, such as apoptosis and necroptosis or, alternatively, tumor necrosis (Chen et al., 2022). Cellular changes due to RT damage manifest within hours post-RT and can persist for days (Chen et al., 2022). While there is no clinically viable way to directly track cellular level changes in vivo, RT-related tumoral changes may manifest radiographically, providing an opportunity to indirectly assess RT dose response (Sun et al., 2016; Yan & Wang, 2021).

Radiomics analyses, which extract quantitative features from imaging data have shown some promise for predicting RT response and decision support (Gillies et al., 2016), with the predominant literature assessing baseline radiomic features from pre- and post-treatment imaging to make predictions (Liu et al., 2017, p. 20). Delta radiomic features that model variations in quantitative tumor imaging features *intra-treatment* have the potential to add further information as to radiation dose response and recurrence, particularly given that sequential images can serve as internal controls for radiographic changes (Cusumano et al., 2021). However, up until recently, intra-fractional radiomic analyses have shown limited potential, largely due to poor quality on-treatment imaging systems, such as cone beam computer tomography (CT), which have low soft tissue contrast and signal-to-noise ratios (van Timmeren et al., 2019).

Recently, magnetic resonance (MR)-guided RT has witnessed a growing utilization and investigation for various malignancies. This novel technology represents an advance for image-guided RT, characterized by standardized, real-time MR images captured at each administered treatment fraction with superior soft tissue contrast and tumor visualization and localization compared to current cone-beam CT. Furthermore, MR-guidance with breath-hold has also been shown to improve image quality (Lee et al., 2024, p. 20). MR-guided RT for lung SBRT, given its standardized dosing regimens and on-treatment imaging protocols, presents an opportunity to systematically investigate intra-fractional radiographic changes and potentially develop longitudinal and reproducible imaging biomarkers of radiation dose response.

Previous work has investigated MR delta radiomics in the setting of pancreatic cancer and discovered that change in histogram skewness, which represents the asymmetry of voxel intensities in relation to mean intensity, correlated with progression free survival (PFS) (Tomaszewski et al., 2021). To our knowledge, there exists no published study of MR delta radiomics in lung cancer, nor have intra-fraction changes been investigated as a measure of dose-response with granular endpoints. We sought to address this knowledge gap by investigating a database of patients receiving MR-guided lung SBRT. In this study, we first determined if MR radiomic feature extraction is feasible and stable for lung SBRT. Then we sought to determine how 1) MR delta radiomics associate with increasing radiation dose delivery and 2) if on-treatment changes are associated with tumor control and patient survival.

## METHODS

### Dataset and Patient Population

We conducted a retrospective study of patients treated with MR-guided lung SBRT at our institution since the inception of our MR-guided RT program in 2019 with database abstraction in October 2022 (Fig. 1). All patients were treated with five RT fractions, delivered every other weekday, per institutional protocol. This study was conducted in accordance with the Declaration of Helsinki guidelines and was approved by the institutional internal review board (IRB) with a waiver of consent. All patients were treated via the ViewRay MRIdian 0.35T system with Stereotactic magnetic resonance (MR)-guided adaptive radiation therapy (SMART) protocol, using techniques previously described(Lee et al., 2024). Patients were included if they were treated to a tumor (primary or metastasis) within the lung and excluded if they did not have available, complete on-treatment imaging, including an MR-simulation scan and 5 pre-treatment scans from fractions one through five. Clinical outcome data including survival status, radiation start date, locoregional control, and distant metastasis status were abstracted by a trained clinical research coordinator in a Research Electronic Data Capture database and verified by a board-certified radiation oncologist (B.H.K.). For patients with multiple primaries treated, analysis was conducted by treatment course. Reporting follows the Transparent Reporting of a Multivariable Prediction Model for Individual Prognosis (TRIPOD) guidelines (Collins et al., 2015).

### Image Acquisition

MR images in the analysis were acquired in DICOM format on the ViewRay MRIdian system at pretreatment simulation (F0) and prior to each subsequent RT fraction (F1-5) with true fast imaging with steady-state precession (TrueFISP) sequence using 144 slices, and voxel size 1.5×1.5×3 mm, with repetition time (TR)=3 ms and echo time (TE)=1.27 ms. Images were acquired in a comfortable, inspiration breath-hold position. Gross tumor volume (GTV) regions of interest (ROIs) were contoured at simulation by the treating radiation oncologist. They were subsequently aligned to the daily MR scans, and if needed, modified, and confirmed daily by treating radiation oncologists at the time of planning via the MR images.

**Radiomics Pipeline**

Radiomic feature extraction was performed by the Pyradiomics package (van Griethuysen et al., 2017) in Python 3.9.1 per Image Biomarker Standardisation Initiative (IBSI) guidelines (Zwanenburg et al., 2020). All DICOM images and structures were converted to NIFTI format and N4 bias field correction was applied before extraction by the SimpleITK (N. J. Tustison et al., 2010) in Python. The intensity values were quantized with bin count 64 and the default Pyradiomics interpolation method B-spline interpolation was used to resample images. Normalization in the Pyradiomics pipeline was modified by dividing the intensity of each image by the corresponding median heart intensity value, which was felt to be a relatively stable measurement that could be acquired from each thoracic MR. This method controls global intensity variation among fractions (Tomaszewski et al., 2021). Median lung intensity value was not considered for normalization because tumors in lung could alter signal intensity. Median heart intensities were extracted for all fractions per patient. A total of 107 radiomic features for every scan were extracted from the following 7 classes: first order statistics, shape, gray level co-occurrence matrix, gray level run length matrix, gray level size zone matrix, neighboring gray tone difference matrix, and gray level dependence matrix. For full list of radiomic features analyzed with definitions, see Table S1.

**Radiomics Stability**

Radiomic feature stability was evaluated both temporally and spatially. Temporal stability was evaluated by comparing feature values acquired at simulation (SIM) and F1, which generally occurred within two weeks of each other, and between which no radiation had been delivered. Spatial stability was evaluated via random erosions and dilations of the GTV contour to simulate small, expected discrepancies between clinicians. Random contouring perturbations were repeated five times to avoid potential bias (Table S1). Stability was assessed via Lin's Concordance Correlation Coefficient (CCC), with a value > 0.90 indicating stability.

Collinearity among features was assessed using Pearson correlation. Feature pairs with a Pearson correlation higher than 0.90 were identified. The absolute mean correlation of each stable feature with all other features was calculated. In each pair, the feature with a higher mean absolute correlation across all other features was dropped to mitigate collinearity effects. For the final set of stable and non-collinear features, we evaluated delta radiomics relative changes from RT fraction to fraction, normalized by the value of the initial fraction. For example, relative feature change between F1 and F2 is (F2-F1)/F1. Delta radiomics ratios from RT start to finish (F5/F1) were also calculated. Features with consistent directional change per fraction of RT dose delivered were identified as potential radiomic indicators of RT-induced cellular change.

**Clinical Endpoints**

Association of delta radiomic slope (F5/F1) for the final stable and non-collinear features and disease outcomes were evaluated for overall survival (OS), progression-free survival (PFS), local-failure free survival (LFFS), and in-field LFFS (iLFFS). iLFFS is defined as >20% increase in longest diameter compared to nadir of disease extent after treatment and/or biopsy proven and/or residual (equal to or greater than pre-treatment) or new Fludeoxyglucose F18 (FDG) uptake on positron emission tomography (PET). Event times were measured from radiation start date to the time of event. Treatment follow-up was per National Comprehensive Cancer Network (NCCN) guidelines with radiographic CT surveillance generally every 3 months following treatment and response were assessed by Response Evaluation Criteria in Solid Tumors (RECIST).

**Statistical Analysis**

Delta radiomic feature trends were investigated by analyzing median relative change between fractions. The analysis of covariance (ANCOVA) was used to assess the differences between patients with and without locoregional failure for relative F5/F1 skewness by base R package in R V.4.3.1. Univariate and multivariate survival analyses were performed for clinical endpoints via the Lifelines package V.0.27.7 in Python. Univariate Cox proportional hazard regression model was developed to evaluate the effect of each delta radiomics ratio (F5/F1) on a particular event. Benjamini-Hochberg correction (Benjamini & Hochberg, 1995) was performed to adjust the p-value for multiple testing correction. For survival analysis, feature selection with Recursive Feature Elimination (RFE) (Guyon et al., 2002) with Lasso Regression was applied for feature selection to control the number of features and avoid overfitting of the multivariate Cox model. The final Cox model included the covariates selected based on the top four feature rankings, representing the importance of feature (Harrell et al., 1996). Given the prior observation of change in Skewness as a predictor of PFS for pancreatic cancer (Tomaszewski et al., 2021), we attempted to validate this particular feature in our cohort. We evaluated this feature for disease outcomes using both the priorly reported cutoff point of F5/F1 of 0.951 (Tomaszewski et al., 2021) as well as with a conditional inference tree that was created within our cohort with respect to LFFS via the partykit package in R. Kaplan-Meier plots with log-rank tests were used to compare, with $p<0.05$ indicating statistical significance.

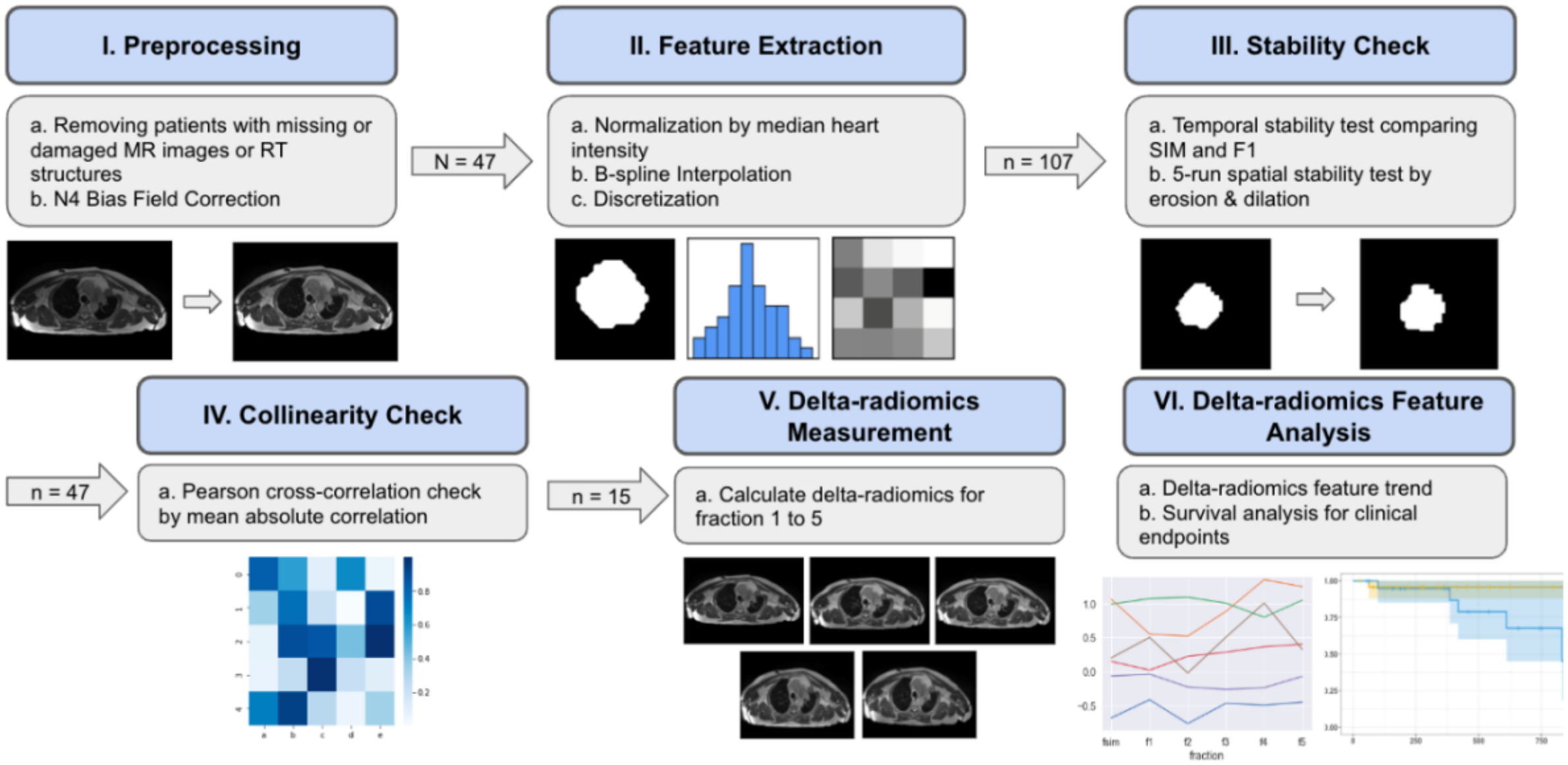

**Figure 1.** Study framework. N indicates number of patients in the study. N indicates number of features after each step. I. Preprocessing. Images before (left) and after (right) show the change after N4 bias field correction. II. Feature Extraction. Three types of features were extracted after preprocessing: shape-based features (left), first-order histogram features (middle), and texture features (right). III. Stability Check. Images before (left) and after (right) show the shape change of tumor after erosion and dilation. IV. Collinearity check. Heatmap for feature Pearson correlation. V. Delta-radiomics Measurement. Delta-radiomics were calculated for fractions 1 (top left), 2 (top middle), 3 (top right), 4 (bottom left), 5 (bottom right). VI. Delta-radiomics Feature Analysis. Relative trend of feature among fractions (left) and Kaplan-Meier for survival analysis (right).

## RESULTS

**Patient characteristics**

A total of 47 MR-guided lung SBRT treatments for 39 patients were included in the study (Table 1). Of these, 22 (47%) treatments were for patients with Stage IV disease. With a median follow-up time of 415 days (range: 32-1178 days), 17 deaths (36%), 25 progression events (53%), 6 locoregional failure events (13%), and specifically 4 in-field locoregional failure events (9%) were observed. The six patients who experienced locoregional failure events each had a single treatment course. Radiation total dose delivered for patients ranged from 40 Gy to 60 Gy in 5 fractions (dose per fraction range: 8 Gy/fx to 12 Gy/fx). One patient had reirradiation to an overlapping area (24 months after prior course). Eight patients received chemotherapy after SBRT.

**Table 1.** Patient and Tumor Characteristics

| **Patient and Treatment Characteristics** | | **N=47 treatments (39 patients)** |
|---|---|---|
| **Age at time of treatment,** median (IQR) | | 67 (61.5-73.5) |
| **Sex** | | |
| | Female | 27 (57%) |
| | Male | 20 (43%) |
| **Smoking status** | | |
| | Former smoker (quit >1 year prior to diagnosis) | 33 (70%) |
| | Never smoker (<100 cigarettes in lifetime) | 10 (21%) |
| | Current smoker (smoking at time of diagnosis or quit <1 year prior) | 3 (6%) |
| | Unspecified | 1 (2%) |
| **Performance status** | | |
| | 0 | 23 (49%) |
| | 1 | 21 (45%) |
| | 2 | 2 (4%) |
| | Unspecified | 1 (2%) |
| **Prior thoracic RT** | | |
| | Yes | 19 (40%) |
| | No | 22 (47%) |
| | Unspecified | 6 (13%) |
| **T stage** | | |
| | T0 | 0 (0%) |
| | T1a | 12 (26%) |
| | T1b | 5 (11%) |
| | T2a | 1 (2%) |
| | T2b | 2 (4%) |
| | T3 | 6 (13%) |
| | T4 | 3 (6%) |
| | Unspecified | 18 (38%) |
| **N stage** | | |
| | N0 | 27 (57%) |
| | N1 | 1 (2%) |
| | N2 | 0 (0%) |
| | N3 | 1 (2%) |
| | Unspecified | 18 (38%) |
| **M stage** | | |
| | M0 | 25 (53%) |
| | M1 | 22 (47%) |
| **Overall stage** | | |
| | I | 15 (32%) |
| | II | 7 (15%) |
| | III | 3 (6%) |
| | IV | 22 (47%) |
| **Recurred tumor** | | |
| | No | 37 (79%) |
| | Yes, reirradiation | 1 (2%) |
| | Yes, not reirradiation | 9 (19%) |
| **Tumor size as radiographically measured at diagnosis, cm (median, IQR)** | | 1.6 (1.3-2.5) |
| **Maximum 3D tumor diameter extracted from GTV, cm (median, IQR)** | | 3.7 (3.0-5.0) |
| **Tumor volume, $cm^3$ (median, IQR)** | | 8.6 (4.3-15.1) |
| **Histology** | | |
| | Metastasis from other site | 15 (32%) |
| | Adenocarcinoma | 14 (30%) |
| | No pathology | 12 (26%) |
| | Mesothelioma | 3 (6%) |
| | Large cell carcinoma | 2 (4%) |
| | Other | 1 (2%) |
| **Radiation total dose delivered (Gy)** | | |
| | 40 | 4 (9%) |
| | 50 | 28 (60%) |
| | 55 | 3 (6%) |
| | 60 | 12 (26%) |
| **Radiation dose per fraction (Gy/fx)** | | |
| | 8 | 4 (9%) |
| | 10 | 28 (60%) |
| | 11 | 3 (6%) |
| | 12 | 12 (26) |
| **Biologically equivalent dose to tumor (Gy10)** | | |
| | 72 | 4 (9%) |
| | 100 | 26 (55%) |
| | 115.5 | 3 (6%) |
| | 132 | 13 (28%) |
| | 300 | 1 (2%) |
| **Reirradiation** | | |
| | Yes | 1 (2%) |
| | No | 46 (98%) |
| **Concurrent chemotherapy** | | |
| | Yes | 0 (0%) |
| | No | 39 (83%) |
| | Unspecified | 8 (17%) |
| **Adjuvant chemotherapy** | | |
| | Yes | 8 (17%) |
| | No | 31 (66%) |
| | Unspecified | 8 (17%) |

Multiple treatments of one patient are considered independent cases. Median of age excluded unspecified patients. IQR: interquartile range, GTV: gross tumor volume. Maximum 3D diameter was extracted via Pyradiomics package using the 3D Feret Diameter. Metastasis from other site included sarcoma, colorectal cancer, melanoma, esophageal cancer, head and neck cancer, and other cancer.

**MR delta radiomics and radiation dose-response**

From 107 features extracted, 49 (46%) demonstrated temporal stability, 57 (53%) demonstrated spatial stability, and 47 (44%) demonstrated both (Table S1). Within the 47 temporally and spatially stable features, 15 non-collinear features were identified (Fig. S1). Stable and non-collinear delta radiomic features demonstrated heterogeneous intra-fraction trajectories across patients (Fig. 2). For most features, there were a mix of increasing and decreasing feature values per fraction (Fig. 2). Notably, Skewness and SurfaceVolumeRatio were the most likely to decrease from F1 to F5 (64%, and 55% of tumors, respectively) across tumors (Fig. 2, Table S2). In contrast, features Coarseness and 90 Percentile were the most likely to increase with radiation dose (62% and 60% of cases, respectively, Table S2). 3D tumor volume (VoxelVolume) was excluded from the final stable features due to collinearity with other variables, but was analyzed separately due to its clinical importance and associations with long-term RT response (Adjogatse et al., 2023) (Table S4 – S8). For all delta radiomic features, across patients, there were substantial portions of patients with all positive (median: +21%, range: +9% to +36%) or all-negative (median: -21%, range: -9% to -30%) relative radiomic value changes for F2 through F5 compared to the F1 baseline (Fig. S2, Table S3). One textural feature, GrayLevelNonUniformity, for which lower values indicate more homogeneity (van Griethuysen et al., 2017) demonstrated consistent negative median change per fraction, and one feature, LargeDependenceHighGrayLevelEmphasis, which indicates concentration of high intensity voxel values, showed a consistent median increase per fraction (Table S4). These two features and Skewness had the largest median absolute changes per fraction on-treatment (Table 2). Skewness demonstrated the largest magnitude of median relative dose-related change from baseline with a range of 22% to 45% per fraction (Table 2).

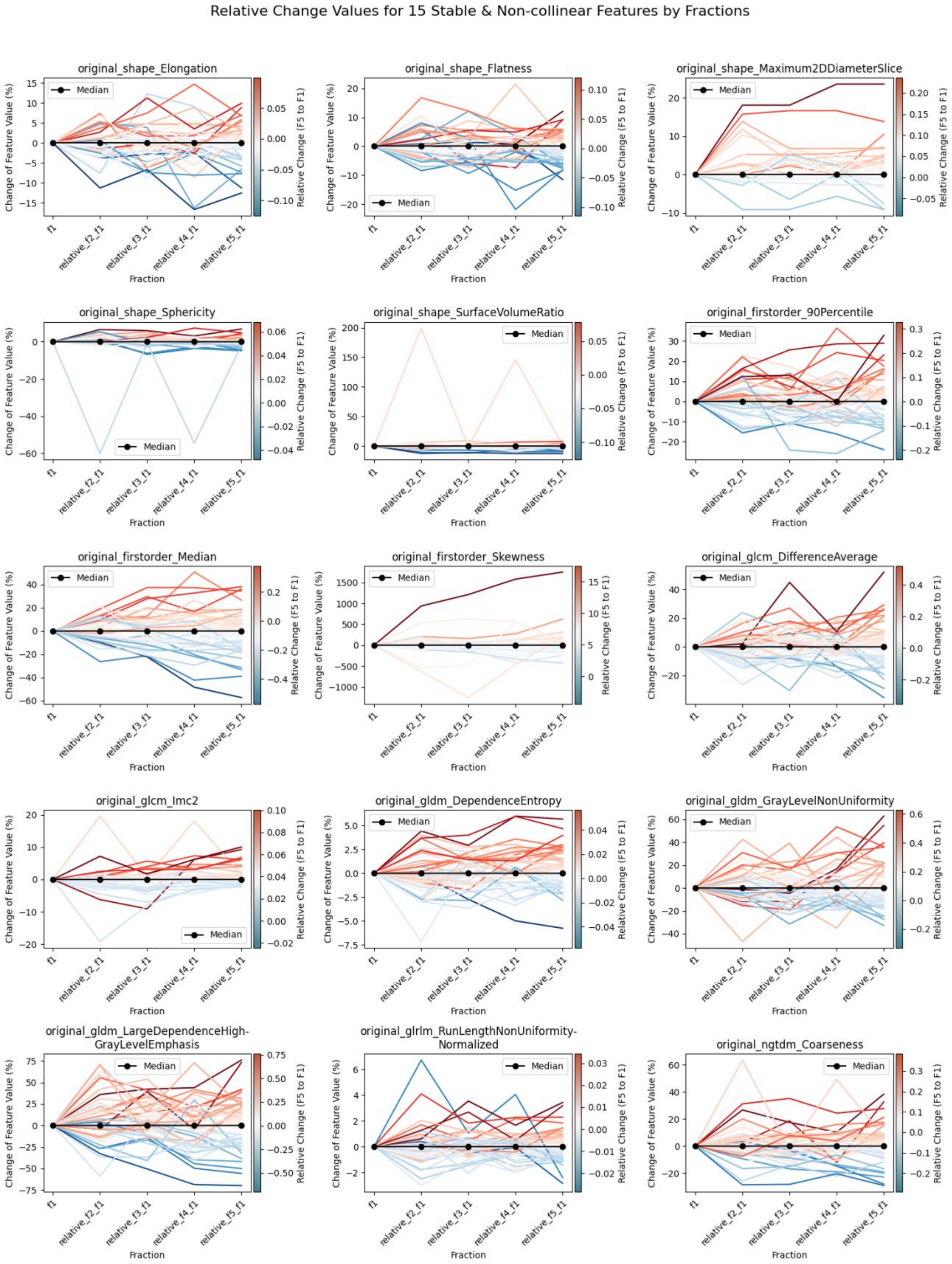
Relative Change Values for 15 Stable & Non-collinear Features by Fractions
original_shape_Elongation
original_shape_Flatness
original_shape_Maximum2DDiameterSlice
original_shape_Sphericity
original_shape_SurfaceVolumeRatio
original_firstorder_90Percentile
original_firstorder_Median
original_firstorder_Skewness
original_glcm_DifferenceAverage
original_glcm_Imc2
original_gldm_DependenceEntropy
original_gldm_GrayLevelNonUniformity
original_gldm_LargeDependenceHigh-GrayLevelEmphasis
original_glrlm_RunLengthNonUniformity-Normalized
original_ngtdm_Coarseness
Median
Change of Feature Value (%)
Relative Change (F5 to F1)
Fraction
f1
relative_f2_f1
relative_f3_f1
relative_f4_f1
relative_f5_f1

**Figure 2.** Delta radiomics trends for stable, non-collinear features across all treatment courses (n=47), showing relative feature changes of feature values across fractions. The dotted black line shows median values of relative value for each fraction. Gradient color legend and color of lines show the value of relative F5 over F1 for each feature.

**Table 2.** Relative median and IQR of absolute change for 15 stable and non-collinear features across fractions 1 to 5 (normalized by fraction 1).

| Feature Name | F2 (%) | F3 (%) | F4 (%) | F5 (%) |
|---|---|---|---|---|
| **Skewness** | 21.79 (10.42, 85.24) | 41.33 (20.49, 72.96) | 40.38 (20.45, 118.55) | 45.41 (22.41, 94.31) |
| **LargeDependenceHighGrayLevelEmphasis** | 16.71 (4.47, 26.98) | 12.50 (5.66, 24.54) | 20.95 (9.12, 31.33) | 18.95 (10.63, 29.83) |
| **GrayLevelNonUniformity** | 8.07 (4.57, 13.24) | 8.76 (5.38, 15.43) | 9.79 (4.53, 15.58) | 11.41 (4.94, 21.54) |
| **DifferenceAverage** | 7.94 (3.42, 11.15) | 6.01 (3.51, 9.35) | 5.62 (2.91, 10.14) | 8.76 (3.59, 18.45) |
| **Median** | 8.10 (3.85, 12.45 | 8.09 (3.15, 13.66) | 9.56 (4.26, 16.59) | 7.71 (2.95, 17.23) |
| **90Percentile** | 5.67 (2.08, 10.45) | 5.13 (2.61, 11.00) | 5.06 (2.38, 10.98) | 7.15 (3.62, 13.42) |
| **Coarseness** | 5.20 (2.51, 9.53) | 5.78 (2.49, 9.76) | 6.56 (3.52, 10.91) | 5.95 (2.72, 15.99) |
| **Flatness** | 2.00 (0.42, 4.94) | 2.70 (1.11, 5.43) | 2.00 (0.75, 4.75) | 2.79 (1.38, 5.69) |
| **Elongation** | 1.58 (0.66, 3.60) | 2.33 (0.59, 4.79) | 1.83 (0.48, 3.39) | 2.10 (0.59, 4.75) |
| **Maximum2DDiameterSlice** | 0.00 (0.00, 2.59) | 1.18 (0.00, 4.62) | 0.00 (0.00, 2.67) | 1.09 (0.00, 3.96) |
| **Imc2** | 0.93 (0.33, 2.11) | 1.04 (0.23, 2.80) | 0.92 (0.35, 3.04) | 1.05 (0.30, 1.91) |
| **DependenceEntropy** | 1.28 (0.46, 2.53) | 1.37 (0.80, 1.84) | 1.38 (0.48, 2.43) | 1.04 (0.67, 2.26) |
| **Sphericity** | 0.61 (0.17, 1.67) | 0.84 (0.33, 2.03) | 0.60 (0.21, 1.62) | 0.95 (0.41, 2.46) |
| **SurfaceVolumeRatio** | 1.05 (0.24, 3.09) | 1.13 (0.50, 2.38) | 0.66 (0.20, 2.30) | 0.79 (0.33, 1.99) |
| **RunLengthNonUniformityNormalized** | 0.72 (0.32, 1.18) | 0.57 (0.36, 1.12) | 0.53 (0.25, 0.95) | 0.75 (0.39, 1.16) |

Median values were calculated by relative feature absolute values. Features were ordered by relative F5 over F1 values, showing intra-fraction fluctuations.

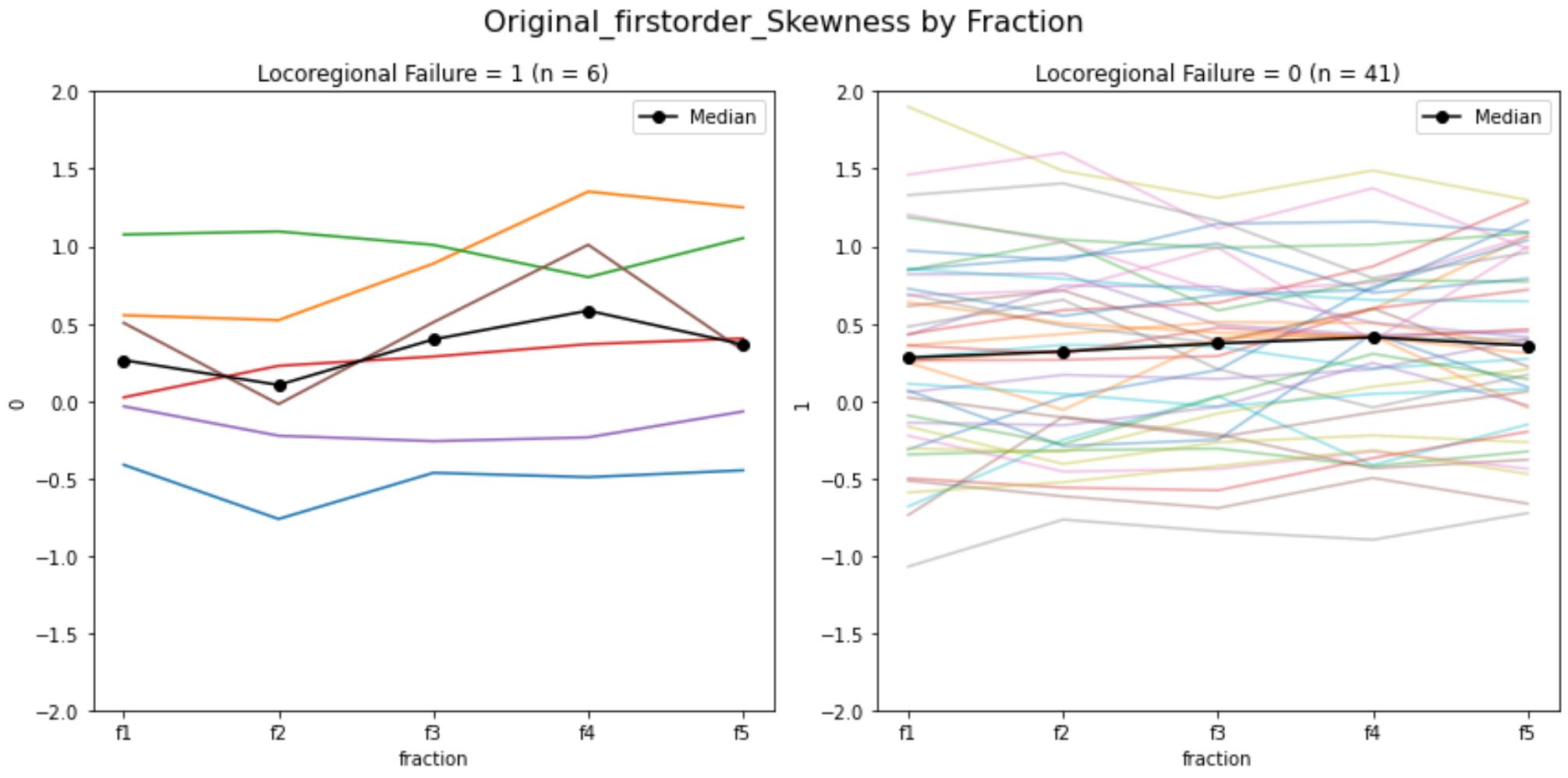

**Figure 3.** Original (top) values of feature skewness from F1 to F5 (normalized by F1) for patients with (n = 6) and without (n = 41) locoregional failure. Medians of the patients of two groups for each fraction were calculated.

### Delta radiomics and disease outcome

For the stable and non-collinear features, univariable survival analysis of four clinical endpoints using feature delta radiomic slope (F5/F1) was examined (Table 3). Initially, we observed significant associations of delta Elongation (*p*=0.006), Flatness (*p*=0.03), and Skewness (p = 0.01) with locoregional failure free survival (LFFS); Imc2 (*p*=0.04) for progression free survival (PFS), and GrayLevelNonUniformity and overall survival (OS) (*p*=0.049). However, after correcting for multiple hypothesis testing, these were no significant associations. It was previously observed that the ratio of feature histogram skewness was statistically significant for PFS of pancreatic cancer (Tomaszewski et al., 2021). In this study, we found Skewness trended towards association with LFFS (*p*=0.01, corrected *p*=0.09), with a greater skewness slope associated with worse outcomes, consistent with prior work. Additionally, ANCOVA analysis demonstrated that F5/F1 Skewness was associated with locoregional failure (*p*=0.01). The relative changes of feature Skewness from F1 to F5 for cases with (n=6) and without (n=41) locoregional failure showed that relative values were sensitive to dose change for both patient groups (Fig. 2). Tumor volume had no observed association with disease outcomes (Table S8).

**Table 3.** Univariable Cox regression analysis for delta radiomic feature relative change (F5/F1) and disease and survival endpoints.

| | **Delta Radiomic Features** | **PFS p-** | **p_bh** | **OS p-** | **p_bh** | **LFFS p-** | **p_bh** | **ILFFS p-** | **p_bh** |
|---|---|---|---|---|---|---|---|---|---|
| **Shape-based** | Elongation | 0.05 | 0.41 | 0.39 | 0.71 | **0.01** | 0.09 | 0.07 | 0.36 |
| | Flatness | 0.28 | 0.71 | 0.59 | 0.71 | **0.03** | 0.15 | 0.07 | 0.36 |
| | Maximum2DDiameterSlice | 0.24 | 0.71 | 0.60 | 0.71 | 0.71 | 0.81 | 0.76 | 0.97 |
| | Sphericity | 0.90 | 0.90 | 0.47 | 0.71 | 0.35 | 0.66 | 0.11 | 0.36 |
| | SurfaceVolumeRatio | 0.79 | 0.90 | 0.97 | 0.97 | 0.48 | 0.66 | 0.15 | 0.36 |
| **First Order** | 90Percentile | 0.40 | 0.78 | 0.86 | 0.92 | 0.48 | 0.66 | 0.86 | 0.97 |
| | Median | 0.85 | 0.90 | 0.40 | 0.71 | 0.76 | 0.81 | 0.92 | 0.97 |
| | Skewness | 0.42 | 0.78 | 0.61 | 0.71 | **0.01** | 0.09 | 0.97 | 0.97 |
| **Gray Level Co-occurrence Matrix (GLCM)** | DifferenceAverage | 0.26 | 0.71 | 0.16 | 0.48 | 0.32 | 0.66 | 0.14 | 0.36 |
| | Imc2 | **0.04** | 0.41 | 0.06 | 0.29 | 0.08 | 0.31 | 0.09 | 0.36 |
| **Gray Level Dependence Matrix (GLDM)** | DependenceEntropy | 0.61 | 0.81 | 0.14 | 0.48 | 0.64 | 0.80 | 0.54 | 0.90 |
| | GrayLevelNonUniformity | 0.65 | 0.81 | **0.049** | 0.29 | 0.94 | 0.94 | 0.96 | 0.97 |
| | LargeDependenceHighGrayLevelEmphasis | 0.49 | 0.81 | 0.24 | 0.61 | 0.48 | 0.66 | 0.69 | 0.97 |
| **Gray Level Run Length Matrix (GLRLM)** | RunLengthNonUniformityNormalized | 0.28 | 0.71 | 0.06 | 0.29 | 0.44 | 0.66 | 0.32 | 0.60 |
| **Neighbouring Gray Tone Difference Matrix (NGTDM)** | Coarseness | 0.65 | 0.81 | 0.54 | 0.71 | 0.43 | 0.66 | 0.22 | 0.47 |

PFS: Progression Free Survival. OS: Overall Survival. LFFS: Locoregional Failure Free Survival. ILFFS: In-field Locoregional Failure Free Survival. P_bh: adjusted p-value after Benjamini/Hochberg correction. P-values <0.05 are bolded.

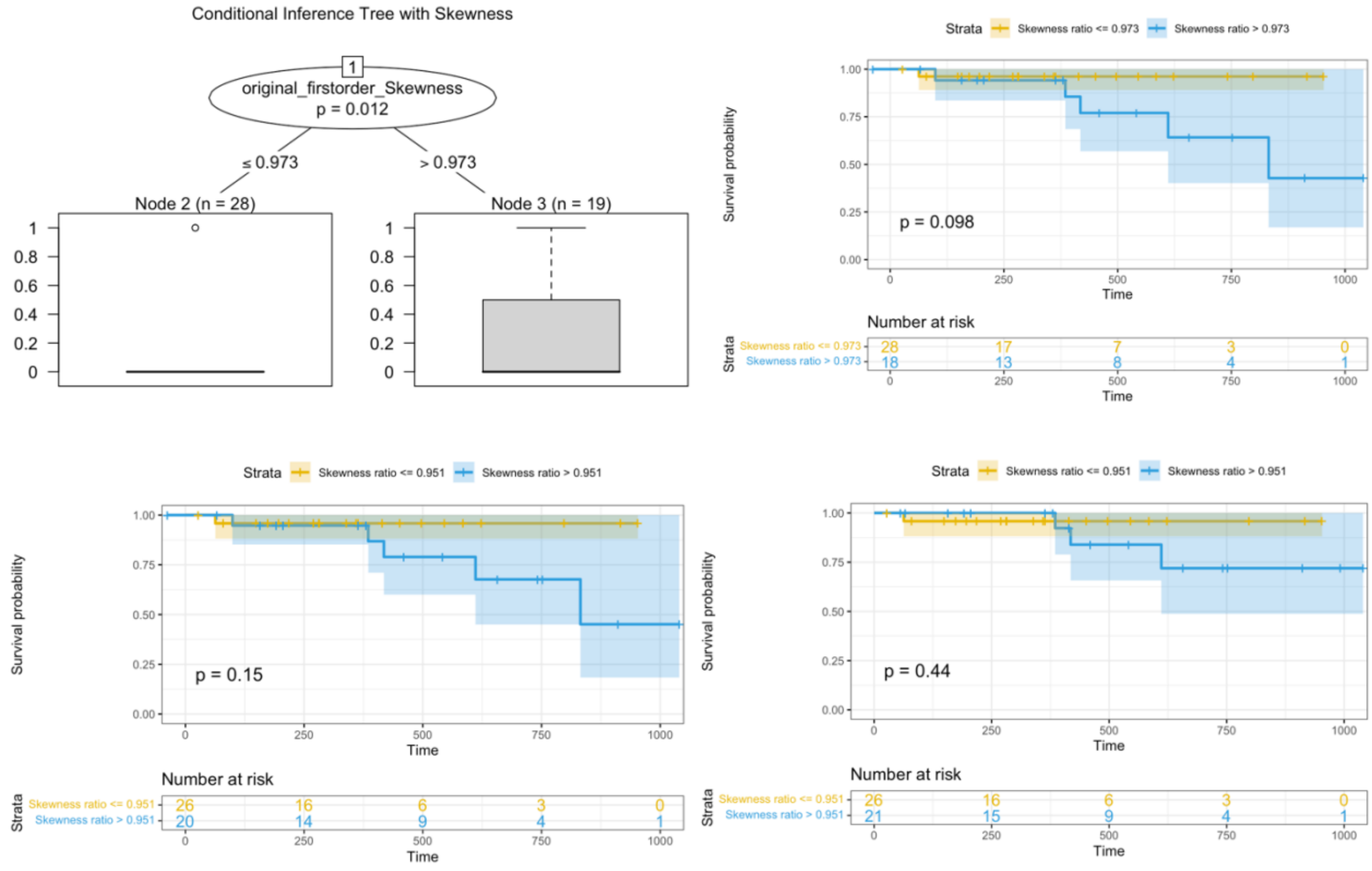

**Figure 4.** Conditional Inference Tree and Kaplan-Meier Plot for Feature Skewness between F5 and F1. Upper Left: Conditional inference tree for LFFS with optimal cutoff point skewness = 0.973. Upper right: Measure the optimal cutoff point for LFFS. Bottom left: Test cutoff point skewness = 0.951 for LFFS. Bottom right: Test cutoff point skewness = 0.951 for ILFFS.

We further investigated the optimal threshold for relative skewness change as a marker for local control. A conditional inference tree using recursive partitioning demonstrated maximal separation to divide patients into low-risk (n=28) and high-risk (n=19) groups of LFFS at an optimal threshold of skewness ratio = 0.973 (Fig. 4), which was very similar to the prior reported threshold for predicting failure in pancreatic cancer (0.951) (Tomaszewski et al., 2021). Survival plots demonstrated numerical separation of curves at each of these thresholds, though these differences did not reach statistical significance.

For multivariable Cox regression with RFE, we found that F5/F1 ratio of feature Elongation, which shows the relationship between the two largest principal components in the GTV shape, was selected to fit the model in all 4 clinical endpoints. Covariates F5/F1 ratio of Elongation, original_glcm_Imc2, and Elongation were all significantly associated with PFS (HR 1.21 [95% CI 1.01-1.45], *p*=0.04; HR 1.2 [95% CI 1.04-1.39], *p*=0.01); HR=1.7 [95% CI 1.10-2.65], *p*=0.02).

**Table 4.** Multivariate Survival Analysis Result.

| **Survival** | **Delta radiomic feature (F5/F1)** | **p-value** | **HR (95% CI)** | **Importance** | **Concordance** |
|---|---|---|---|---|---|
| **PFS** | | | | | 0.68 |
| | Elongation | **0.04** | 1.21 (1.01 – 1.45) | 1 | |
| | Flatness | 0.15 | 0.86 (-.71 – 1.05) | 2 | |
| | 90Percentile | 0.06 | 0.97 (0.93 – 1.00) | 4 | |
| | Imc2 | **0.01** | 1.20 (1.04 – 1.39) | 3 | |
| **OS** | | | | | 0.63 |
| | Elongation | 0.43 | 1.09 (0.88 – 1.37) | 2 | |
| | Flatness | 0.63 | 0.94 (0.75 – 1.19) | 3 | |
| | SurfaceVolumeRatio | 0.81 | 0.98 (0.81 – 1.18) | 4 | |
| | DependenceEntropy | 0.16 | 1.20 (0.93 – 1.57) | 1 | |
| **LFFS** | | | | | 0.9 |
| | Elongation | **0.02** | 1.70 (1.10 – 2.65) | 2 | |
| | Flatness | 0.18 | 0.74 (0.48 – 1.15) | 3 | |
| | DependenceEntropy | 0.82 | 1.08 (0.56 – 2.06) | 4 | |
| | RunLengthNonUniformityNormalized | 0.32 | 1.67 (0.61 – 4.57) | 1 | |
| **ILFFS** | | | | | 0.85 |
| | Elongation | 0.07 | 1.25 (0.99 – 1.59) | 2 | |
| | Median | 0.91 | 1.00 (0.93 – 1.09) | 3 | |
| | Imc2 | 0.08 | 1.53 (0.95 – 2.46) | 1 | |
| | DependenceEntropy | 0.38 | 0.65 (0.24 – 1.71) | 4 | |

HR: hazard ratio. CI: confidence interval. Importance: covariate importance ranking by RFE. Concordance: the amount of agreement by Cox model. A p-value less than 0.05 was statistically significant, which is bolded.

## DISCUSSION

Radiation-induced tumor intra- and extra-cellular response may manifest radiographically, and the ability to practically measure and evaluate these changes would significantly advance personalized radiotherapy and management. MR-guided RT, with its superior intra-fraction imaging quality and standardization, coupled with rigorous radiomics methodologies, presents a promising pathway to capture radiographic manifestations of radiation dose effect in a clinical setting. To our knowledge, this is the first study to systematically characterize MR radiomic intra-treatment changes for lung tumors and builds on the nascent literature investigating MR-guided RT delta-radiomics since this technology reached the clinic several years ago. In this proof-of-principle study, we found that delta radiomic feature extraction can be feasible, stable, and reproducible on a low-field, 0.35T MRI-guided linear accelerator. We investigated 15 reproducible features and uncovered distinct trajectories across patients as radiation dose is delivered. We discovered several radiomic features that change consistently with radiation dose delivered and that are associated with tumor control and survival, including evidence that voxel Skewness may be a cross-cancer marker of radiation response.

Remarkably, many features extracted were unstable to temporal and spatial perturbations, emphasizing the importance of rigorous stability testing even when evaluating intra-patient changes in a setting with standardized imaging acquisition parameters. Several features, most notably Skewness and LargeDependenceHighGrayLevelEmphasis showed consistent median changes across treatment fractions, suggesting that these changes may reflect cumulative dose effects. Radiomic features that are sensitive to dose change, like Skewness, could have utility in response prediction due to their quantifiable change. Additionally, associations observed between skewness and LFFS in this study and prior work (Tomaszewski et al., 2021) suggest its potential as a predictive biomarker for treatment efficacy, potentially across cancer types. We hypothesize that voxel Skewness may reflect intratumoral heterogeneity and that fluctuating heterogeneity through treatment may be indicative of a local response. Further work is ongoing to investigate how Skewness and other textural radiomic features specifically correlate with cellular heterogeneity. For multivariate survival models, radiomics features were more highly concordant with local failure outcomes than overall progression or survival, suggesting that these changes reflect local phenomena. Given the early stage of data collection and aggregation for MR-guided delta radiomics, and the small, heterogeneous patient population in this study, further validation will be necessary to confirm these findings, which should be treated as exploratory and hypothesis generating. We expect these findings and the methodologies in this study to form a basis for further validation studies in larger cohorts across malignancy types as volumes and experience with MR-guided RT systems increase.

Our work builds upon the extremely nascent literature in RT delta radiomics by addressing several crucial aspects. Several older studies had evaluated CT-based delta radiomic features with limited promise (Li, Kim, Balagurunathan, Liu, et al., 2017; Li, Kim, Balagurunathan, Qi, et al., 2017). Only one study, in the setting of pancreatic SBRT, has been published evaluating MR-guided RT delta radiomics (Tomaszewski et al., 2021), which demonstrated the potential of MR delta radiomic features, particularly Skewness, in predicting RT response. Our study expands the investigations of MR delta radiomics to the setting of lung cancer, and the first to systematically evaluate radiomic feature trajectories after both temporal and spatial stability testing.

This study has several limitations, the predominant being the small sample size and single-institution source, which may lead to selection bias. It's important to note that this study was limited to one single MRI sequence, and the total dose was not accounted for in our analysis. Our analysis is necessarily simplistic in looking mainly at linear relationships between inter-fraction radiomic values and their association with radiation delivery. Additionally, our dataset was too small to compare total dose differences. The relationship between RT-induced cellular change and radiographic manifestation may be more complex and non-linear. Larger, multi-institutional datasets will be needed to employ more complex statistical and machine learning modeling that may uncover these relationships. Interpretation of radiomic associations with outcomes such as survival and overall progression should be done with caution, given the heterogeneity of our cohort, which included a large proportion of Stage IV patients. While heterogeneous radiomic feature responses through treatment were clearly observed, further work will be needed to validate specific delta radiomic patterns and signatures that consistently (and reproducibly) track with RT-induced cellular change. Other specific methodologies should be explored as well. In the future, adding image filters before feature extraction to increase the number of features that may highlight more detailed and comprehensive image properties and capture variation of dose response more accurately (Demircioğlu, 2022; Shur et al., 2021). Since we found skewness can potentially track dose response, nonspatial filters like taking the square or exponential could be used to adjust the sensitivity of skewness-related radiomics features to intensity values (Shur et al., 2021). Deep learning may provide an attractive means of predicting outcomes based on delta radiomic features (Dudas et al., 2023), though these models generally require much larger training cohorts than what is presently available.
In conclusion, our study demonstrates a feasible and stable delta radiomics pipeline for MR-guided lung SBRT that can serve as a framework for future cancer radiomics investigations. We find that MR-delta radiomic features change heterogeneously with distinct trajectories across tumors, and that several may be indicative of RT-related tumor cellular change with increasing dose delivered, as well as risk of recurrence. Further work is needed to build on these findings to determine if delta radiomics can ultimately serve as real-time, clinical biomarkers to guide personalized radiation regimens and therapeutic strategies.

## SUPPLEMENTARY MATERIALS

**Table S1.** Stability Check for Extracted Features.

| Feature class | Feature Name | Temporal | Spatial | Spatial | Spatial | Spatial | Spatial | Stable |
|---|---|---|---|---|---|---|---|---|
| Shape-based | Elongation | 0.9813 | 0.9777 | 0.9725 | 0.9675 | 0.9696 | 0.9760 | 1 |
| | Flatness | 0.9811 | 0.9728 | 0.9631 | 0.9776 | 0.9675 | 0.9708 | 1 |
| | LeastAxisLength | 0.9923 | 0.9870 | 0.9865 | 0.9869 | 0.9826 | 0.9890 | 1 |
| | MajorAxisLength | 0.9961 | 0.9972 | 0.9975 | 0.9980 | 0.9972 | 0.9982 | 1 |
| | Maximum2DDiameterColumn | 0.9951 | 0.9972 | 0.9972 | 0.9977 | 0.9958 | 0.9968 | 1 |
| | Maximum2DDiameterRow | 0.9939 | 0.9978 | 0.9972 | 0.9982 | 0.9968 | 0.9974 | 1 |
| | Maximum2DDiameterSlice | 0.9960 | 0.9858 | 0.9827 | 0.9818 | 0.9826 | 0.9886 | 1 |
| | Maximum3DDiameter | 0.9949 | 0.9958 | 0.9958 | 0.9974 | 0.9954 | 0.9973 | 1 |
| | MeshVolume | 0.9732 | 0.9963 | 0.9983 | 0.9968 | 0.9967 | 0.9980 | 1 |
| | MinorAxisLength | 0.9928 | 0.9897 | 0.9908 | 0.9901 | 0.9890 | 0.9908 | 1 |
| | Sphericity | 0.9775 | 0.9457 | 0.9609 | 0.9704 | 0.9592 | 0.9718 | 1 |
| | SurfaceArea | 0.9913 | 0.9931 | 0.9956 | 0.9945 | 0.9934 | 0.9952 | 1 |
| | SurfaceVolumeRatio | 0.9849 | 0.9843 | 0.9821 | 0.9850 | 0.9735 | 0.9890 | 1 |
| | VoxelVolume | 0.9732 | 0.9963 | 0.9983 | 0.9968 | 0.9967 | 0.9979 | 1 |
| First Order Statistics | 10Percentile | 0.9031 | 0.9703 | 0.9736 | 0.9714 | 0.9520 | 0.9731 | 1 |
| | 90Percentile | 0.9426 | 0.9621 | 0.9879 | 0.9816 | 0.9812 | 0.9537 | 1 |
| | original_firstorder_Energy | 0.9445 | 0.9951 | 0.9982 | 0.9953 | 0.9976 | 0.9967 | 1 |
| | original_firstorder_Entropy | 0.7501 | 0.8738 | 0.9038 | 0.8766 | 0.8739 | 0.9095 | 0 |
| | original_firstorder_InterquartileRange | 0.7927 | 0.9033 | 0.9213 | 0.9218 | 0.8877 | 0.8808 | 0 |
| | original_firstorder_Kurtosis | 0.7952 | 0.8784 | 0.9462 | 0.8881 | 0.8881 | 0.9134 | 0 |
| | original_firstorder_Maximum | 0.8871 | 0.9174 | 0.9629 | 0.9372 | 0.9334 | 0.9625 | 0 |
| | original_firstorder_MeanAbsoluteDeviation | 0.8829 | 0.8991 | 0.9414 | 0.9304 | 0.9084 | 0.8858 | 0 |
| | original_firstorder_Mean | 0.9420 | 0.9861 | 0.9872 | 0.9868 | 0.9860 | 0.9884 | 1 |
| | original_firstorder_Median | 0.9391 | 0.9811 | 0.9791 | 0.9801 | 0.9807 | 0.9908 | 1 |
| | original_firstorder_Minimum | 0.8826 | 0.9478 | 0.9410 | 0.9092 | 0.9094 | 0.9043 | 0 |
| | original_firstorder_Range | 0.8900 | 0.9035 | 0.9507 | 0.9209 | 0.9172 | 0.9491 | 0 |
| | original_firstorder_RobustMeanAbsoluteDeviation | 0.8396 | 0.8969 | 0.9278 | 0.9133 | 0.8852 | 0.8781 | 0 |
| | original_firstorder_RootMeanSquared | 0.9457 | 0.9862 | 0.9899 | 0.9877 | 0.9889 | 0.9842 | 1 |
| | original_firstorder_Skewness | 0.9096 | 0.9429 | 0.9584 | 0.9443 | 0.9530 | 0.9653 | 1 |
| | original_firstorder_TotalEnergy | 0.9446 | 0.9953 | 0.9983 | 0.9958 | 0.9977 | 0.9972 | 1 |
| | original_firstorder_Uniformity | 0.8003 | 0.8778 | 0.8987 | 0.8881 | 0.8550 | 0.9068 | 0 |
| | original_firstorder_Variance | 0.8942 | 0.8753 | 0.9435 | 0.9214 | 0.9091 | 0.8525 | 0 |
| Gray Level Co-occurrence Matrix (GLCM) | original_glcm_Autocorrelation | 0.7523 | 0.8599 | 0.9200 | 0.8976 | 0.8918 | 0.8967 | 0 |
| | original_glcm_ClusterProminence | 0.6033 | 0.8507 | 0.9167 | 0.8991 | 0.8819 | 0.9072 | 0 |
| | original_glcm_ClusterShade | 0.8660 | 0.9179 | 0.9571 | 0.9607 | 0.9484 | 0.9400 | 0 |
| | original_glcm_ClusterTendency | 0.7467 | 0.8782 | 0.9192 | 0.9023 | 0.8958 | 0.9290 | 0 |
| | original_glcm_Contrast | 0.8969 | 0.8920 | 0.9461 | 0.9199 | 0.9596 | 0.9666 | 0 |
| | original_glcm_Correlation | 0.8294 | 0.7924 | 0.8309 | 0.8586 | 0.8179 | 0.8558 | 0 |
| | original_glcm_DifferenceAverage | 0.9040 | 0.9016 | 0.9476 | 0.9239 | 0.9515 | 0.9619 | 1 |
| | original_glcm_DifferenceEntropy | 0.8785 | 0.9198 | 0.9492 | 0.9269 | 0.9367 | 0.9557 | 0 |
| | original_glcm_DifferenceVariance | 0.8669 | 0.9106 | 0.9547 | 0.9235 | 0.9685 | 0.9725 | 0 |
| | original_glcm_Id | 0.9055 | 0.9078 | 0.9477 | 0.9283 | 0.9305 | 0.9532 | 1 |
| | original_glcm_Idm | 0.9045 | 0.9063 | 0.9461 | 0.9286 | 0.9250 | 0.9516 | 1 |
| | original_glcm_Idmn | 0.9001 | 0.8937 | 0.9461 | 0.9197 | 0.9575 | 0.9652 | 0 |
| | original_glcm_Idn | 0.9048 | 0.9045 | 0.9479 | 0.9253 | 0.9485 | 0.9603 | 1 |
| | original_glcm_Imc1 | 0.9708 | 0.9536 | 0.9791 | 0.9712 | 0.9714 | 0.9798 | 1 |
| | original_glcm_Imc2 | 0.9322 | 0.9720 | 0.9705 | 0.9746 | 0.9632 | 0.9641 | 1 |
| | original_glcm_InverseVariance | 0.9040 | 0.9170 | 0.9417 | 0.9392 | 0.9255 | 0.9521 | 1 |
| | original_glcm_JointAverage | 0.7961 | 0.8520 | 0.9196 | 0.8961 | 0.8914 | 0.9109 | 0 |
| | original_glcm_JointEnergy | 0.8224 | 0.8657 | 0.8831 | 0.8938 | 0.8261 | 0.9063 | 0 |
| | original_glcm_JointEntropy | 0.8735 | 0.9261 | 0.9332 | 0.9322 | 0.8927 | 0.9451 | 0 |
| | original_glcm_MCC | 0.7986 | 0.8062 | 0.7944 | 0.7845 | 0.7287 | 0.7507 | 0 |
| | original_glcm_MaximumProbability | 0.6978 | 0.7752 | 0.8068 | 0.7616 | 0.7201 | 0.8526 | 0 |
| | original_glcm_SumAverage | 0.7961 | 0.8520 | 0.9196 | 0.8961 | 0.8914 | 0.9109 | 0 |
| | original_glcm_SumEntropy | 0.7556 | 0.8929 | 0.8934 | 0.8750 | 0.8610 | 0.8963 | 0 |
| | original_glcm_SumSquares | 0.7936 | 0.8979 | 0.9391 | 0.9155 | 0.9281 | 0.9504 | 0 |
| Gray Level Dependence Matrix (GLDM) | original_gldm_DependenceEntropy | 0.9243 | 0.9453 | 0.9606 | 0.9479 | 0.9408 | 0.9499 | 1 |
| | original_gldm_DependenceNonUniformity | 0.9935 | 0.9971 | 0.9969 | 0.9977 | 0.9950 | 0.9974 | 1 |
| | original_gldm_DependenceNonUniformityNormalized | 0.8885 | 0.8920 | 0.9589 | 0.9131 | 0.9467 | 0.9555 | 0 |
| | original_gldm_DependenceVariance | 0.8977 | 0.9116 | 0.9557 | 0.9214 | 0.9380 | 0.9667 | 0 |
| | original_gldm_GrayLevelNonUniformity | 0.9204 | 0.9935 | 0.9988 | 0.9924 | 0.9984 | 0.9977 | 1 |
| | original_gldm_GrayLevelVariance | 0.7522 | 0.8829 | 0.9284 | 0.9009 | 0.9126 | 0.9388 | 0 |
| | original_gldm_HighGrayLevelEmphasis | 0.7337 | 0.8487 | 0.9142 | 0.8960 | 0.8770 | 0.8910 | 0 |
| | original_gldm_LargeDependenceEmphasis | 0.9086 | 0.9211 | 0.9568 | 0.9323 | 0.9392 | 0.9660 | 1 |
| | original_gldm_LargeDependenceHighGrayLevelEmphasis | 0.9267 | 0.9729 | 0.9736 | 0.9585 | 0.9721 | 0.9637 | 1 |
| | original_gldm_LargeDependenceLowGrayLevelEmphasis | 0.6482 | 0.3483 | 0.8375 | 0.8737 | 0.6572 | 0.7287 | 0 |
| | original_gldm_LowGrayLevelEmphasis | 0.6632 | 0.6958 | 0.8971 | 0.8953 | 0.7791 | 0.9006 | 0 |
| | original_gldm_SmallDependenceEmphasis | 0.8945 | 0.9165 | 0.9566 | 0.9292 | 0.9430 | 0.9523 | 0 |
| | original_gldm_SmallDependenceHighGrayLevelEmphasis | 0.6720 | 0.8153 | 0.9107 | 0.8885 | 0.8632 | 0.8874 | 0 |
| | original_gldm_SmallDependenceLowGrayLevelEmphasis | 0.7088 | 0.8116 | 0.9288 | 0.8974 | 0.8777 | 0.9408 | 0 |
| Gray Level Run Length Matrix (GLRLM) | original_glrlm_GrayLevelNonUniformity | 0.9330 | 0.9951 | 0.9988 | 0.9944 | 0.9984 | 0.9983 | 1 |
| | original_glrlm_GrayLevelNonUniformityNormalized | 0.7987 | 0.8829 | 0.9032 | 0.8923 | 0.8596 | 0.9084 | 0 |
| | original_glrlm_GrayLevelVariance | 0.7461 | 0.8819 | 0.9276 | 0.8991 | 0.9121 | 0.9377 | 0 |
| | original_glrlm_HighGrayLevelRunEmphasis | 0.7308 | 0.8485 | 0.9142 | 0.8955 | 0.8760 | 0.8900 | 0 |
| | original_glrlm_LongRunEmphasis | 0.9119 | 0.9234 | 0.9578 | 0.9357 | 0.9396 | 0.9642 | 1 |
| | original_glrlm_LongRunHighGrayLevelEmphasis | 0.7591 | 0.8655 | 0.9209 | 0.9014 | 0.8884 | 0.8977 | 0 |
| | original_glrlm_LongRunLowGrayLevelEmphasis | 0.6519 | 0.6530 | 0.8919 | 0.8947 | 0.7617 | 0.8851 | 0 |
| | original_glrlm_LowGrayLevelRunEmphasis | 0.6618 | 0.7056 | 0.9000 | 0.8955 | 0.7852 | 0.9040 | 0 |
| | original_glrlm_RunEntropy | 0.7049 | 0.8731 | 0.8952 | 0.8695 | 0.8645 | 0.8929 | 0 |
| | original_glrlm_RunLengthNonUniformity | 0.9839 | 0.9973 | 0.9980 | 0.9978 | 0.9963 | 0.9978 | 1 |
| | original_glrlm_RunLengthNonUniformityNormalized | 0.9085 | 0.9196 | 0.9562 | 0.9323 | 0.9369 | 0.9595 | 1 |
| | original_glrlm_RunPercentage | 0.9113 | 0.9210 | 0.9572 | 0.9339 | 0.9386 | 0.9619 | 1 |
| | original_glrlm_RunVariance | 0.9138 | 0.9263 | 0.9591 | 0.9378 | 0.9416 | 0.9667 | 1 |
| | original_glrlm_ShortRunEmphasis | 0.9087 | 0.9199 | 0.9560 | 0.9328 | 0.9368 | 0.9601 | 1 |
| | original_glrlm_ShortRunHighGrayLevelEmphasis | 0.7248 | 0.8449 | 0.9129 | 0.8944 | 0.8736 | 0.8887 | 0 |
| | original_glrlm_ShortRunLowGrayLevelEmphasis | 0.6640 | 0.7176 | 0.9022 | 0.8955 | 0.7904 | 0.9076 | 0 |
| Gray Level Size Zone Matrix (GLSZM) | original_glszm_GrayLevelNonUniformity | 0.9975 | 0.9963 | 0.9983 | 0.9950 | 0.9976 | 0.9986 | 1 |
| | original_glszm_GrayLevelNonUniformityNormalized | 0.7062 | 0.8902 | 0.9139 | 0.8736 | 0.8824 | 0.9122 | 0 |
| | original_glszm_GrayLevelVariance | 0.6493 | 0.8611 | 0.9101 | 0.8578 | 0.9043 | 0.9270 | 0 |
| | original_glszm_HighGrayLevelZoneEmphasis | 0.6942 | 0.8465 | 0.9122 | 0.8914 | 0.8647 | 0.8819 | 0 |
| | original_glszm_LargeAreaEmphasis | 0.4863 | 0.8028 | 0.9988 | 0.8258 | 0.9983 | 0.9644 | 0 |
| | original_glszm_LargeAreaHighGrayLevelEmphasis | 0.5395 | 0.8922 | 0.9994 | 0.8945 | 0.9988 | 0.9830 | 0 |
| | original_glszm_LargeAreaLowGrayLevelEmphasis | 0.4370 | 0.6871 | 0.9748 | 0.7453 | 0.9697 | 0.9135 | 0 |
| | original_glszm_LowGrayLevelZoneEmphasis | 0.6382 | 0.7771 | 0.9153 | 0.8853 | 0.8508 | 0.9168 | 0 |
| | original_glszm_SizeZoneNonUniformity | 0.9921 | 0.9951 | 0.9959 | 0.9940 | 0.9943 | 0.9964 | 1 |
| | original_glszm_SizeZoneNonUniformityNormalized | 0.7951 | 0.8921 | 0.9467 | 0.9059 | 0.9349 | 0.9268 | 0 |
| | original_glszm_SmallAreaEmphasis | 0.7944 | 0.9041 | 0.9447 | 0.9097 | 0.9305 | 0.9251 | 0 |
| | original_glszm_SmallAreaHighGrayLevelEmphasis | 0.6514 | 0.8295 | 0.9055 | 0.8856 | 0.8523 | 0.8782 | 0 |
| | original_glszm_SmallAreaLowGrayLevelEmphasis | 0.6378 | 0.7842 | 0.9166 | 0.8813 | 0.8821 | 0.9206 | 0 |
| | original_glszm_ZoneEntropy | 0.8993 | 0.9456 | 0.9457 | 0.9368 | 0.9326 | 0.9512 | 0 |
| | original_glszm_ZonePercentage | 0.9122 | 0.9254 | 0.9575 | 0.9372 | 0.9426 | 0.9588 | 1 |
| | original_glszm_ZoneVariance | 0.4848 | 0.8017 | 0.9988 | 0.8250 | 0.9983 | 0.9642 | 0 |
| Neighbouring Gray Tone Differenc | original_ngtdm_Busyness | 0.9443 | 0.9774 | 0.9695 | 0.9772 | 0.9629 | 0.9911 | 1 |
| | original_ngtdm_Coarseness | 0.9756 | 0.9681 | 0.9809 | 0.9717 | 0.9549 | 0.9886 | 1 |
| e Matrix (NGTDM) | original_ngtdm_Complexity | 0.6648 | 0.8356 | 0.8901 | 0.8364 | 0.9004 | 0.9341 | 0 |
| | original_ngtdm_Contrast | 0.9166 | 0.8891 | 0.9502 | 0.9375 | 0.9584 | 0.9621 | 0 |
| | original_ngtdm_Strength | 0.9487 | 0.9790 | 0.9874 | 0.9694 | 0.9726 | 0.9940 | 1 |

Temporal stability measures the difference between SIM and F1 fractions. Spatial stability was conducted 5 times by erosion and dilation. Lin's Concordance Correlation Coefficient (CCC) > 0.90 was used to assess stability. 47 features with both temporal and spatial stability were labeled as 1 and used for feature analysis.

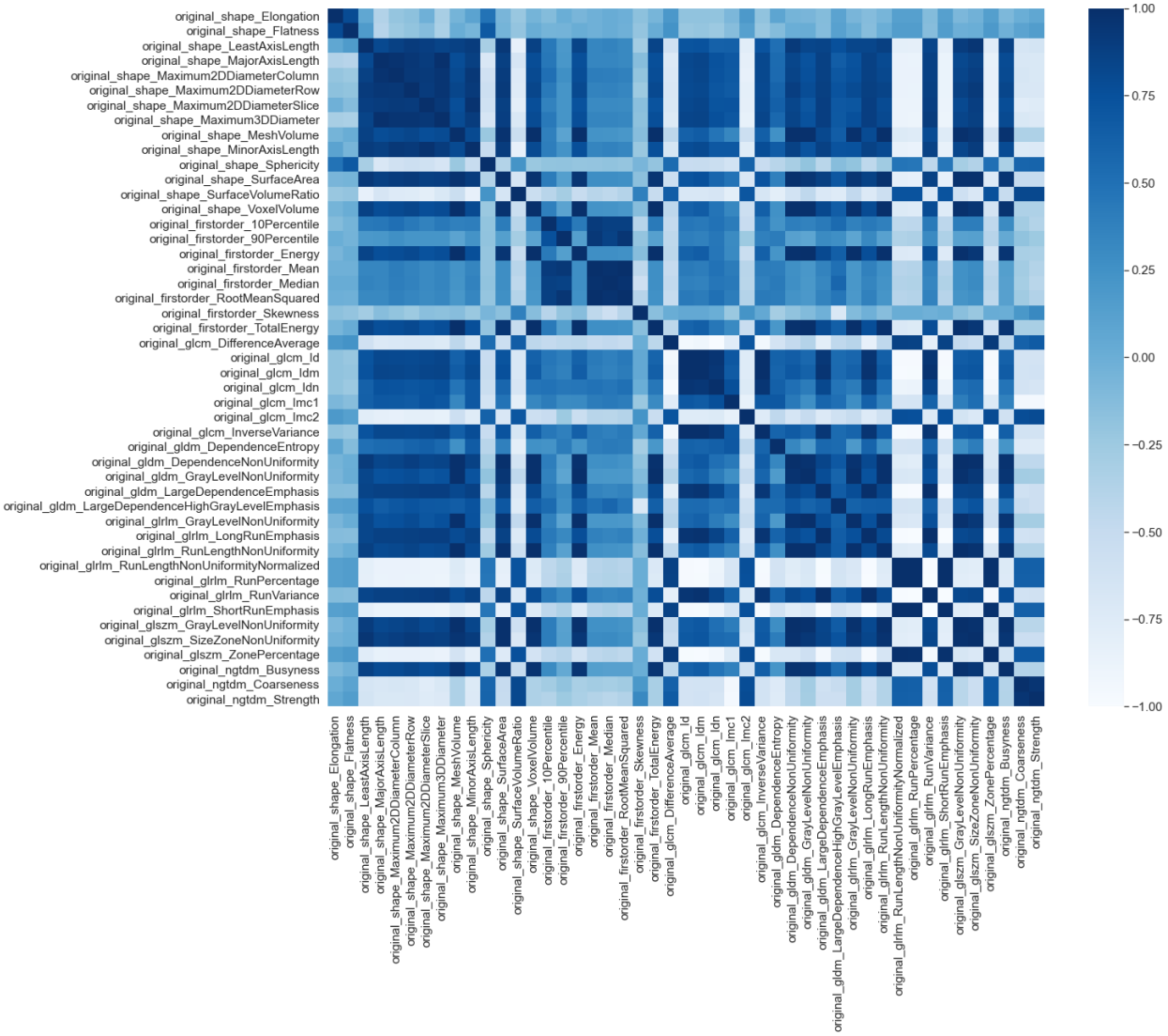

**Figure S1.** Collinearity Check for Stable Features. Pearson cross correlation was performed on 47 stable F1 features. A 47 x 47 Pearson correlation matrix was made, and correlation threshold

was set to 0.9. Each pair of features with a correlation higher than 0.9 was chosen. For each feature pair, one with higher mean absolute correlation with all other features was dropped.

**Table S2.** Percentage of positive, negative, or no relative change from F1 to F5 in 15 stable and non-collinear features across fractions. In cases of volume or certain shape-based features, no change could occur if the clinician did not modify the tumor segmentation from prior treatment fraction.

| **Feature Name** | **Positive (%)** | **Negative (%)** | **No Change (%)** |
|---|---|---|---|
| **original_ngtdm_Coarseness** | 61.70 | 38.30 | 0.00 |
| **original_firstorder_90Percentile** | 59.57 | 40.43 | 0.00 |
| **original_gldm_DependenceEntropy** | 57.45 | 42.55 | 0.00 |
| **original_gldm_LargeDependenceHighGrayLevelEmphasis** | 57.45 | 42.55 | 0.00 |
| **original_glcm_DifferenceAverage** | 53.19 | 46.81 | 0.00 |
| **original_shape_Sphericity** | 53.19 | 44.68 | 2.13 |
| **original_firstorder_Median** | 51.06 | 48.94 | 0.00 |
| **original_shape_Elongation** | 51.06 | 46.81 | 2.13 |
| **original_shape_Flatness** | 48.94 | 48.94 | 2.13 |
| **original_glcm_Imc2** | 46.81 | 53.19 | 0.00 |
| **original_gldm_GrayLevelNonUniformity** | 46.81 | 53.19 | 0.00 |
| **original_glrlm_RunLengthNonUniformityNormalized** | 46.81 | 53.19 | 0.00 |
| **original_shape_SurfaceVolumeRatio** | 42.55 | 55.32 | 2.13 |
| **original_shape_Maximum2DDiameterSlice** | 42.55 | 23.40 | 34.04 |
| **original_firstorder_Skewness** | 36.17 | 63.83 | 0.00 |

**Table S3.** Percentage of patients with consistent positive (relative change > 0) or negative (relative change < 0) trends in 15 stable and non-collinear features across fractions.

| Feature | Cases with all relative change > 0 (%) | Cases with all relative change < 0 (%) | Total Consistent Cases (%) | Other Cases (%) |
|---|---|---|---|---|
| original_firstorder_90Percentile | 34.04 | 25.53 | 59.57 | 40.43 |
| original_firstorder_Median | 29.79 | 29.79 | 59.57 | 40.43 |
| original_gldm_DependenceEntropy | 31.91 | 21.28 | 53.19 | 46.81 |
| original_gldm_LargeDependenceHighGrayLevelEmphasis | 36.17 | 14.89 | 51.06 | 48.94 |
| original_glcm_Imc2 | 25.53 | 23.40 | 48.94 | 51.06 |
| original_firstorder_Skewness | 19.15 | 25.53 | 44.68 | 55.32 |
| original_gldm_GrayLevelNonUniformity | 21.28 | 23.40 | 44.68 | 55.32 |
| original_glcm_DifferenceAverage | 23.40 | 19.15 | 42.55 | 57.45 |
| original_shape_Flatness | 17.02 | 23.40 | 40.43 | 59.57 |
| original_ngtdm_Coarseness | 25.53 | 14.89 | 40.43 | 59.57 |
| original_glrlm_RunLengthNonUniformityNormalized | 19.15 | 19.15 | 38.30 | 61.70 |
| original_shape_SurfaceVolumeRatio | 8.51 | 25.53 | 34.04 | 65.96 |
| original_shape_Elongation | 19.15 | 14.89 | 34.04 | 65.96 |
| original_shape_Sphericity | 19.15 | 10.64 | 29.79 | 70.21 |
| original_shape_Maximum2DDiameterSlice | 19.15 | 8.51 | 27.66 | 72.34 |
| Median | 21.28 | 21.28 | 42.55 | 57.45 |

Total consistent cases showed percent of cases with either consistent positive or negative change. Other Cases column showed the percent of cases demonstrating mixed relative changes. Median was calculated for each column across all features.

**Table S4.** Relative median change for 15 stable and non-collinear features across fractions 1 to 5 (normalized by fraction 1).

| **Feature Name** | **F2 (%)** | **F3 (%)** | **F4 (%)** | **F5 (%)** |
|---|---|---|---|---|
| original_gldm_LargeDependenceHighGrayLevelEmphasis | 4.12 | 4.68 | 0.55 | 3.40 |
| original_firstorder_90Percentile | 1.73 | -0.68 | 1.34 | 2.52 |
| original_ngtdm_Coarseness | -0.08 | -1.30 | 1.88 | 0.81 |
| original_glcm_DifferenceAverage | 0.16 | 2.81 | -1.50 | 0.65 |
| original_gldm_DependenceEntropy | -0.17 | 0.60 | 0.32 | 0.58 |
| original_firstorder_Median | 2.62 | 0.08 | 0.00 | 0.31 |
| original_shape_Elongation | 0.00 | 0.00 | 0.00 | 0.00 |
| original_shape_Sphericity | 0.14 | 0.07 | 0.00 | 0.00 |
| original_shape_Flatness | 0.00 | 0.00 | 0.00 | 0.00 |
| original_shape_Maximum2DDiameterSlice | 0.00 | 0.00 | 0.00 | 0.00 |
| original_shape_SurfaceVolumeRatio | -0.10 | -0.27 | -0.11 | 0.00 |
| original_glcm_Imc2 | 0.01 | 0.07 | 0.23 | -0.06 |
| original_glrlm_RunLengthNonUniformityNormalized | 0.08 | 0.20 | 0.17 | -0.07 |
| original_gldm_GrayLevelNonUniformity | -1.41 | -1.78 | -2.93 | -1.98 |
| original_firstorder_Skewness | 1.83 | -11.03 | -14.82 | -8.47 |

Median values were calculated by raw relative feature values. Features were ordered by relative F5 over F1 values, showing consistent directional change.

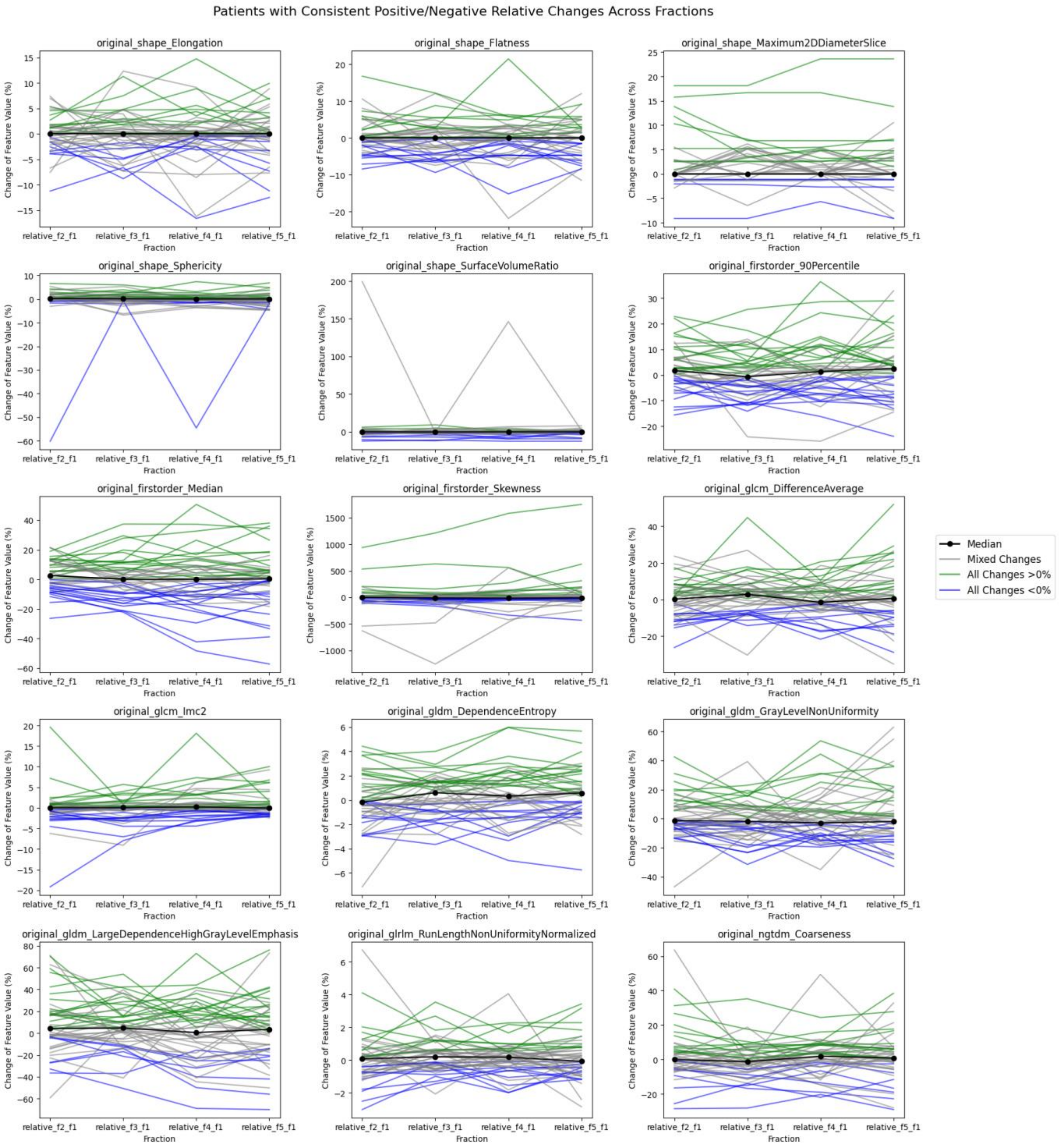

**Figure S2.** Percentage of patients with consistent relative changes in stable and non-collinear features across fractions. The color of each line represents the proportion of patients with either entirely positive, entirely negative, or mixed relative changes. The dotted black line shows median values of relative value for each fraction.

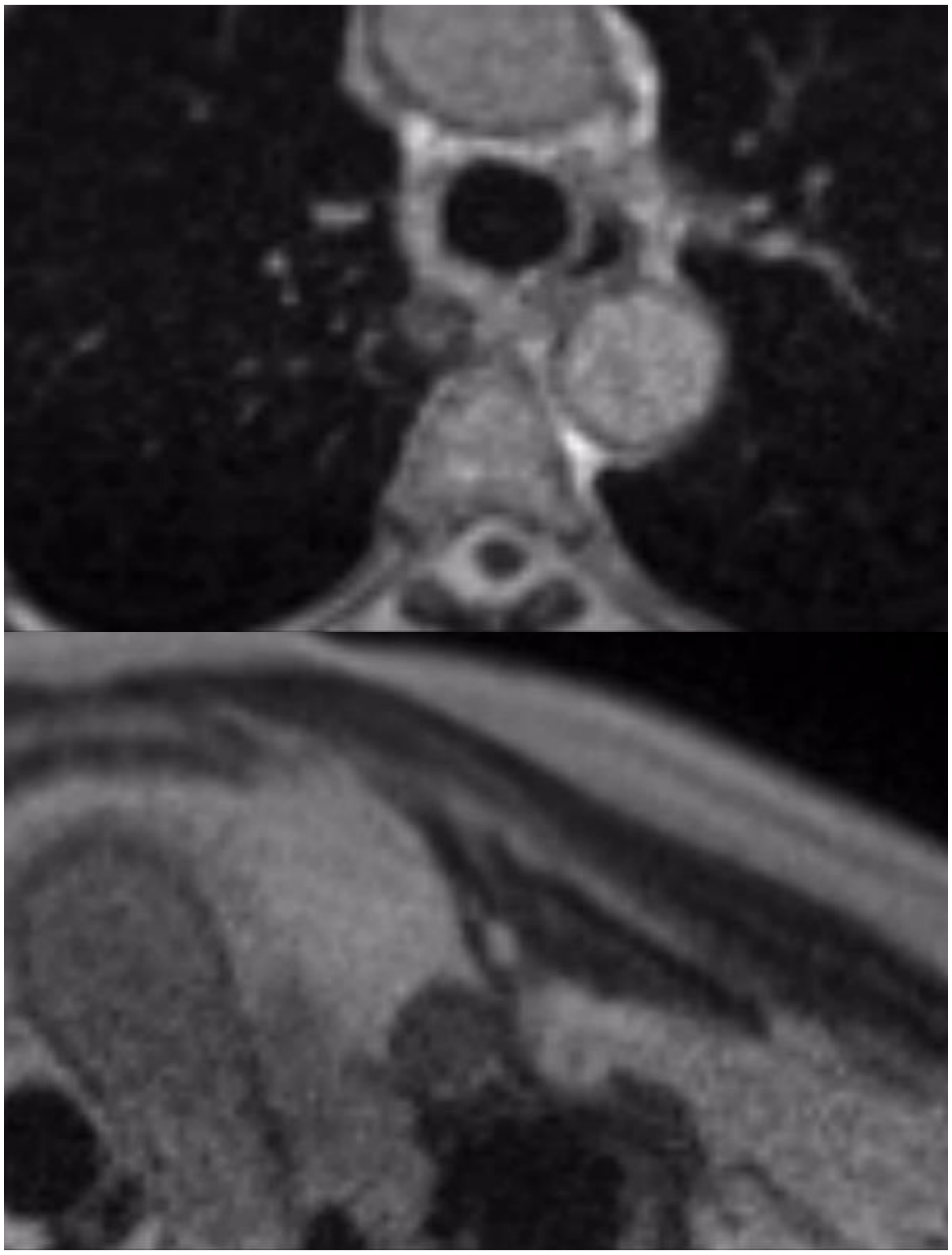

**Figure S3.** Patient with the largest Skewness change from F1 to F5 (top). Patient with the smallest Skewness change from F1 to F5 (bottom).

**Supplementary Text.** Though 3D volume feature VoxelVolume was excluded from feature analysis because of collinearity, it was analyzed because it is associated with RT response (Adjogatse et al., 2023). It showed a tendency to increase with each additional radiation fraction delivered since 66% cases had positive relative change from F1 to F5 across tumors. The percentage of patients exhibiting consistent positive changes across fractions was higher than those exhibiting negative changes (Figure S4). However, it had no large median absolute changes per fraction (Table S5), no consistent pattern of percentage of patients with positive/negative trends (Table S6), and no consistent median increase/decrease (Table S7). It also has no significant association with all clinical endpoints (Table S8).

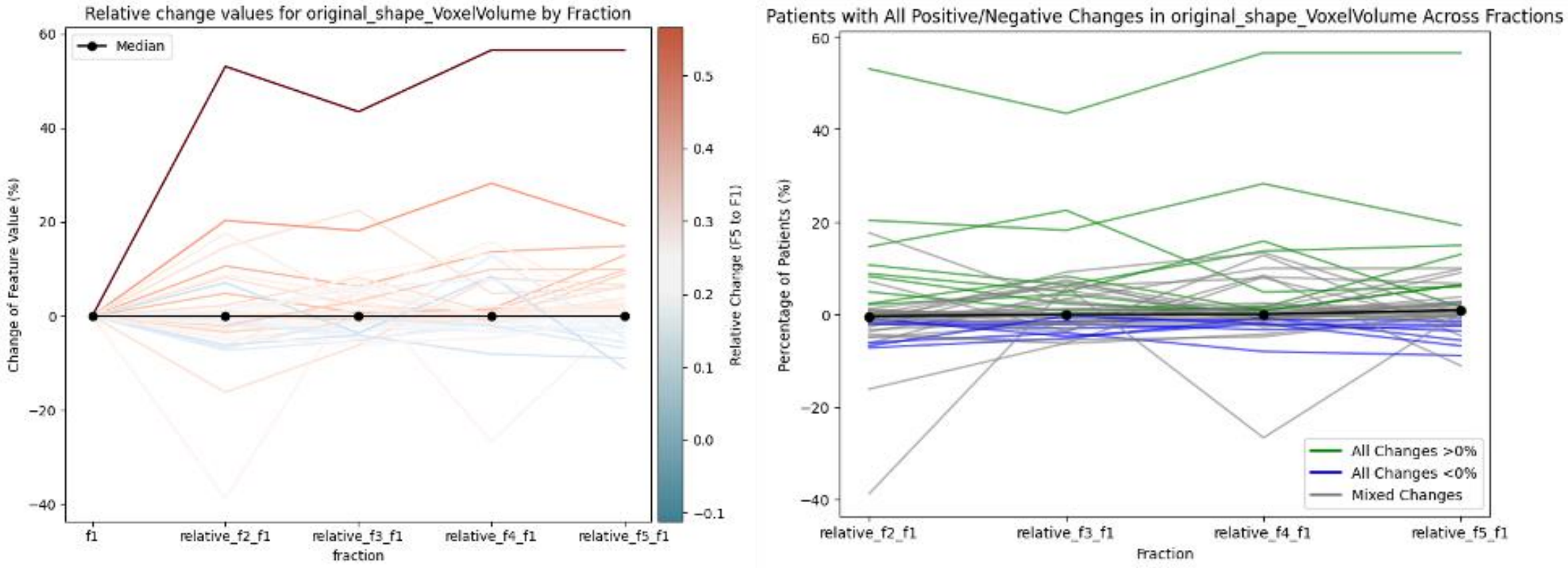

**Figure S4.** Delta radiomics trends for feature VoxelVolume across all treatment courses (n=47), showing relative feature changes of feature values across fractions (left). Gradient color legend and color of lines show the value of relative F5 over F1 for each feature. Percentage of patients with consistent relative changes in feature VoxelVolume across fractions (right). The color of each line represents the proportion of patients with either entirely positive, entirely negative, or mixed relative changes. For both plots, the dotted black line shows median values of relative value for each fraction.

**Table S5.** Relative median and IQR of absolute change for feature VoxelVolume across fractions 1 to 5 (normalized by fraction 1) (top).

| Feature Name | F2 (%) | F3 (%) | F4 (%) | F5 (%) |
|---|---|---|---|---|
| VoxelVolume | 2.10 (0.8, 6.59) | 3.12 (1.13, 5.95) | 1.66 (0.74, 6.35) | 2.23 (0.88, 6.33) |

Median values were calculated by relative feature absolute values. Features were ordered by relative F5 over F1 values, showing intra-fraction fluctuations.

**Table S6.** Percentage of patients with consistent positive (relative change > 0) or negative (relative change < 0) trends in VoxelVolume across fractions.

| Feature | Cases with all relative change > 0 (%) | Cases with all relative change < 0 (%) | Total Consistent Cases (%) | Other Cases (%) |
|---|---|---|---|---|
| VoxelVolume | 19.15 | 14.89 | 34.04 | 65.96 |

**Table S7.** Relative median change for feature VoxelVolume across fractions 1 to 5 (normalized by fraction 1).

| Feature Name | F2 (%) | F3 (%) | F4 (%) | F5 (%) |
|---|---|---|---|---|
| VoxelVolume | -0.46 | 0.00 | 0.00 | 0.88 |

**Table S8.** Univariable Cox regression analysis for delta radiomic feature relative change (F5/F1) and disease and survival endpoints. p_bh was ignored since it's separate univariate analysis only for one feature so p = p_bh.

| | Delta Radiomic Features | PFS p-value | p_bh | OS p-value | p_bh | LFFS p-value | p_bh | ILFFS p-valueva | p_bh |
|---|---|---|---|---|---|---|---|---|---|
| **Shape-based** | VoxelVolume | 0.94 | - | 0.65 | - | 0.65 | - | 0.61 | - |